\documentclass[twoside,12pt]{article}
\usepackage{epsfig}
\usepackage{graphicx}% Include figurefiles
\usepackage{psfrag}

\def\Journal#1#2#3#4{{#1} {#2} (#4) #3 }

\def\NPA{{\em Nucl. Phys.} A}

\def\PRL{\em Phys. Rev. Lett.}

\def\PREP{\em Phys. Rep.}

\def\PRC{{\em Phys. Rev.} C}

\def\RMP{{\em Rev. Mod. Phys.}}

\def\PPNP{{\em Prog.\ Part.\ Nucl.\ Phys.}}

\newcommand{\be}{\begin{equation}}
\newcommand{\ee}{\end{equation}}
\newcommand{\bea}{\begin{eqnarray}}
\newcommand{\eea}{\end{eqnarray}}

\begin{document}

\title{ \vspace{1cm} 
On the possibility to measure nuclear matrix elements of neutrinoless double beta decay in charge-exchange reactions.
}
\author{Vadim Rodin and Amand Faessler
\\
Institut f\"{u}r Theoretische Physik, Universit\"{a}t
T\"{u}bingen, D-72076 T\"{u}bingen, Germany}

\maketitle

\begin{abstract}

As shown in Ref.~\cite{Rod09}, the Fermi nuclear matrix element $M^{0\nu}_F$ of neutrinoless double beta ($0\nu\beta\beta$) decay can be reconstructed if one is able to measure the isospin-forbidden single Fermi transition matrix element from the ground state of the final nucleus to the isobaric analog state (IAS) of the initial nucleus, for instance by means of charge-exchange reactions of the $(n,p)$-type. Here, simple estimates for $^{82}$Se are made which show that indeed the tiny cross section $\sigma_{np}(0_f^+\to IAS)$ is dominated by the admixture of the double IAS in the ground state of the final nucleus provided that the isospin mixing
%between the IAS and the $0^+$ states of normal isospin in the intermediate nucleus 
is weak and can be treated perturbatively. A measurement of such a cross section would definitely be a very difficult task, but it can advance a lot our knowledge of the $0\nu\beta\beta$ nuclear matrix element.

\end{abstract}

Neutrino is the only known spin-$\frac12$ %1/2 
fermion which may be truly neutral, i.e., identical with its own antiparticle. In such a case one speaks about Majorana neutrino, to be contrasted with Dirac neutrino which is different from its antiparticle~\cite{Kay89}. 
Majorana neutrinos naturally appear in many extensions of the standard model (see, e.g.,~\cite{Moh04}). 

A study of nuclear neutrinoless double beta ($0\nu\beta\beta$) decay~~$^A_Z {\mathrm{El}}_N \longrightarrow \ _{Z+2}^{\phantom{+2} A} {\mathrm{El}}_{N-2} + 2e^-$ 
offers the only feasible way to test the charge-conjugation property of neutrinos. 
The existence of $0\nu\beta\beta$ decay would immediately prove neutrino to be a Majorana particle~\cite{vogelbook,AEE07}. The decay also allows to probe the absolute neutrino masses at the level of tens of meV. 

The next generation of $0\nu\beta\beta$-decay experiments (see, e.g., Ref.~\cite{AEE07}  for a recent review) has a great discovery potential. 
Accurate knowledge of the relevant nuclear matrix elements $M^{0\nu}$ will be crucial to reliably deduce the effective Majorana mass from the future experimental data.

\begin{figure}[tb]
\begin{center}
\begin{minipage}[t]{16.5 cm}
%8 cm}
\begin{center}
\includegraphics[scale=.5]{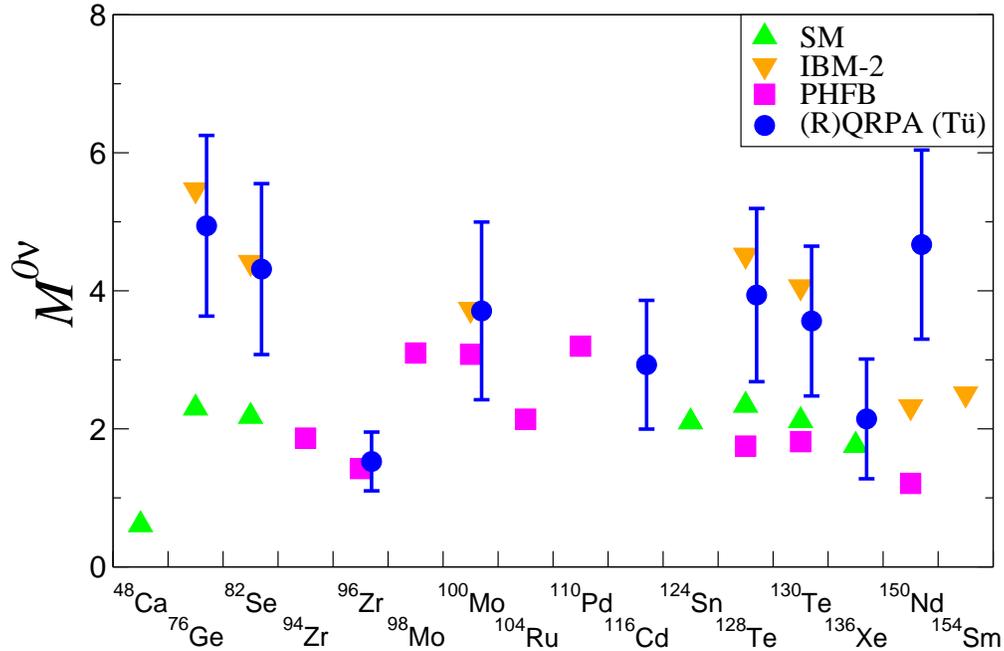} 
\end{center}
\end{minipage}
\begin{minipage}[t]{16.5 cm}
\caption{(Color online) Neutrinoless double beta decay transition matrix elements $M^{0\nu}$ calculated in different approaches: the quasiparticle random phase approximation (QRPA)~\cite{Rod08}, the nuclear shell model (SM)~\cite{Poves}, the projected Hartree-Fock-Bogoliubov method (PHFB)~\cite{Rath} and the Interacting Boson Model (IBM-2)~\cite{Iachello}. The error bars in the QRPA results are calculated from the highest and the lowest values of $M^{0\nu}$ obtained in the calculations with three different single-particle basis sets, %two forces (Bonn CD and Argonne V18)
two different axial vector coupling constants $g_A = 1.25$ and $g_A = 1.00$ (the quenched value), and two different treatments of short range correlations (Jastrow-like
%~\cite{Miller} 
and the UCOM). 
%~\cite{Feld}The radius parameter is as in this whole work $r_{0} = 1.2$ fm.
\label{M0nu}}
\end{minipage}
\end{center}
\end{figure}

Several theoretical approaches have been used to evaluate $M^{0\nu}$.
The present world status of the results on $M^{0\nu}$ for the light neutrino mass mechanism is shown in Fig.~\ref{M0nu} (adopted from Ref.~\cite{esc10}).
Here, we are not going into a detailed discussion of advantages and disadvantages of different models which can be found in Ref.~\cite{esc10}. One only notices that 
the calculated NSM $M^{0\nu}$ scatter substantially, up to a factor of 2.
%Even more striking is the difference in the Fermi contribution to the total $M^{0\nu}$ which can be up to a factor of 5 larger in the QRPA calculations than in the NSM ones.

In such a situation, 
it would be extremely important to find a possibility to determine $M^{0\nu}$ experimentally.
There have been attempts to reconstruct the nuclear amplitude of two-neutrino $\beta\beta$ decay 
from partial one-leg transition amplitudes to the intermediate $1^+$ states measured in charge-exchange reactions~\cite{frekers}. 
However, such a procedure can consistently determine $M^{2\nu}$ only if a transition via a single intermediate 
$1^+$ state dominates $M^{2\nu}$, %(the so-called single-state dominance),
since relative phases of different contributions cannot be measured. 
Trying the same way to reconstruct $M^{0\nu}$ seems even more hopeless, since  
many intermediate states of different multipolarities (with a rather moderate contribution of the $1^+$ states)
are virtually populated in the $0\nu\beta\beta$ decay due to a large momentum of the virtual neutrino.

\begin{figure}[tb]
\begin{center}
\begin{minipage}[t]{16.5 cm}
\psfrag{aa1}{\Large $\hat T^{-}$}\psfrag{bb1}{\Large $\hat T^{+}$}
\psfrag{bb2}{\Large $\hat V_C$}
\includegraphics[scale=.5]{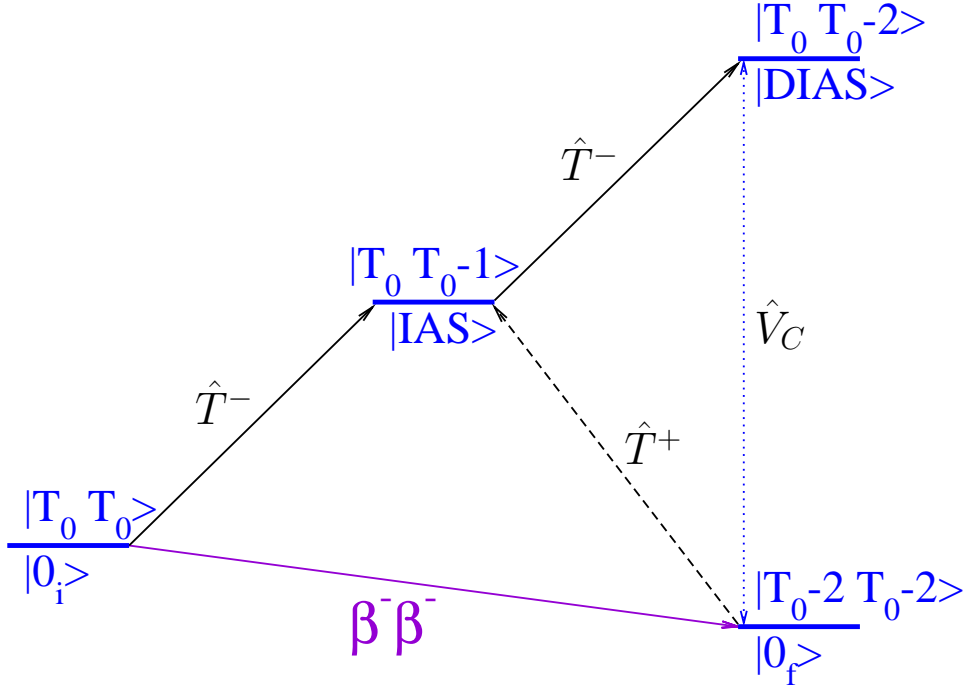} 
\end{minipage}
\begin{minipage}[t]{16.5 cm}
\caption{(Color online) Double Fermi transition from the ground state $| 0_i \rangle$ of the initial nucleus would go exclusively to its DIAS in the final nucleus if the isospin 
%SU(II) 
symmetry were exact. The isotensor part of the Coulomb interaction admixes this ideal DIAS to the g.s. $| 0_f \rangle$ of the final nucleus. Because of the mixing, the single Fermi transition matrix element $\langle IAS | \hat T^{+} | 0_f \rangle$ becomes non-vanishing, and could be measured, e.g., by charge-exchange reactions.\label{DIAS}}
\end{minipage}
\end{center}
\end{figure}

An alternative proposal was put forward in a recent work~\cite{Rod09}. It exploits
the similarity between the Fermi part of the neutrino potential in $0\nu\beta\beta$ decay and the radial dependence of the two-body Coulomb interaction. The latter is well-known to be the leading source of the isospin breaking in nuclei~\cite{auer72}. As shown in Ref.~\cite{Rod09}, the Fermi part $M_{F}^{0\nu}$ of the total matrix element $M^{0\nu}$ can be related to the Coulomb mixing matrix element between the ideal double isobaric analog state (DIAS) of the ground state (g.s.) of the initial nucleus and the g.s. of the final nucleus. This ideal DIAS would be an exact nuclear state if the isospin symmetry were exact. As a result of the Coulomb mixing,
the single Fermi transition matrix element $\langle 0_f | \hat T^{-} | IAS \rangle$ between the isobaric analog state (IAS) of the g.s. of the initial nucleus and the g.s. of the final nucleus becomes non-zero as is illustrated in Fig.~\ref{DIAS}. 
Thus, having measured this single Fermi transition matrix element, e.g., by charge-exchange reactions, the $0\nu\beta\beta$-decay matrix element $M_{F}^{0\nu}$ can be reconstructed. 

The master relation, derived in Ref.~\cite{Rod09} by making use of the isospin 
symmetry of strong interaction, represents 
the matrix element $M_{F}^{0\nu}$ in the form of an energy-weighted double Fermi transition matrix element:
\be
M^{0\nu}_F = - \frac{2}{e^2} 
\sum_s \bar\omega_s \langle 0_f | \hat T^{-} |0^+_s \rangle  \langle 0^+_s | \hat T^{-} |0_i\rangle.
\label{MFtot}
\ee
Here, $\hat T^{-}=\sum_{a}\tau_a^{-}$ is the isospin lowering operator,
the sum runs over all $0^+$ states of the intermediate $(N-1,Z+1)$ isobaric nucleus, 
$\bar\omega_s=E_s-(E_{0_i}+E_{0_f})/2$
is the excitation energy of the $s$'th intermediate state relative to the mean energy of g.s. of the initial and final nucleus.  

As argued in Ref.~\cite{Rod09}, the expression (\ref{MFtot}) must be dominated
by the amplitude of the double Fermi transition via the IAS of the initial g.s. into the final g.s.:
\begin{equation}
M^{0\nu}_F \approx - \frac{2}{e^2}\,\bar\omega_{IAS} 
\langle 0_f | \hat T^{-} |IAS \rangle  \langle IAS | \hat T^{-} |0_i\rangle .
\label{MFappr}
\end{equation}
Here, the second Fermi transition amplitude is non-vanishing due to an admixture of the ideal DIAS 
$|DIAS\rangle=\frac{ (\hat T^{-})^2}{\sqrt{4T_0(2T_0-1)}} |0_i^+\rangle$
in the g.s. of the final nucleus: 
$\langle 0_f |\hat T^{-}| IAS\rangle = \langle 0_f | DIAS\rangle \langle DIAS |\hat T^{-}| IAS\rangle$, $T_0=(N-Z)/2$ is isospin of the g.s. of the initial nucleus. 

Therefore, the total $M^{0\nu}_F$ can be reconstructed according to Eq.~(\ref{MFappr}), 
if one is able to measure the $\Delta T=2$ isospin-forbidden matrix element 
$\langle IAS | \hat T^{+} | 0_f \rangle$, for instance in charge-exchange reactions of the $(n,p)$-type.
Using recent QRPA calculation results for $M^{0\nu}_F$, this matrix element 
can roughly be estimated as 
$\langle IAS | \hat T^{+} | 0_f \rangle \sim 0.005$,  
i.e. about thousand times smaller than the first-leg m.e. $\langle IAS | \hat T^{-} | 0_i \rangle \approx \sqrt{N-Z}$. 
This strong suppression of $\langle IAS | \hat T^{+} | 0_f \rangle$ reflects the smallness of the isospin mixing effects in nuclei. 

The IAS has been observed as a prominent and extremely narrow resonance, and its various features have well been studied 
by means of $(p,n)$, ($^3$He,$t$) and other charge-exchange reactions on the mother nucleus. In this case the reaction cross section at the zero scattering angle can be shown to be proportional to a large Fermi matrix element $\langle IAS | \hat T^{-} | 0_i \rangle \approx \sqrt{2T_0}$~\cite{tad87}.
Extraction of a strongly suppressed matrix element $\langle IAS | \hat T^{+} | 0_f \rangle$ from the tiny cross section of the $(n,p)$-type charge-exchange reactions on the final nucleus might only be possible if there is a proportionality similar to the $(p,n)$ channel. Therefore, a realistic analysis of the corresponding reaction mechanism is needed.
% to assess the feasibility of such a proportionality.

First of all, it is easy to argue that isospin of the projectile may not be larger than $T=1/2$. Indeed, the main components of the wave functions $| 0_f \rangle$ and $| IAS \rangle$ have total isospin different by two units. Therefore, already for a projectile with isospin $T=1$ there exists a common entrance and exit isospin channel arising from recoupling of the isospin $T=1$ of the projectile with the main components of the wave functions of the target and daughter nuclei. In such a case extraction of the information about small isospin impurities from the corresponding reaction cross section seems barely possible.

Thus, the only feasible probes must be of isospin $T=1/2$ ($(n,p)$, ($t,^3$He),\dots reactions). However, even for these probes it is still not guaranteed that the reaction cross section is proportional to a strongly suppressed matrix element $\langle IAS | \hat T^{+} | 0_f \rangle$ since the other isospin impurities in the wave functions $| 0_f \rangle$ and $| IAS \rangle$ may have larger effect on the reaction cross section. In the following we shall assess the case of the $(n,p)$ reaction at the zero scattering angle. We also restrict ourselves to the case of a single, well-isolated, IAS as it appears in rather light nuclei like $^{48}$Ca, neglecting therefore the spreading of the IAS that becomes important in heavier nuclei. In such a case the Coulomb mixing can be treated perturbatively that significantly simplifies the consideration.

The g.s. of the initial nucleus has the largest isospin projection $T_z=T_0$ %=(N-Z)/2 
among all the nuclei involved,
% in the problem, 
therefore it can be considered as a state of the pure total isospin $T=T_0$: $ | 0_i^+ \rangle \equiv |T_0\, T_0\rangle_i $. However, small isospin admixtures induced by the Coulomb mixing should be taken into account in the wave functions of the IAS and the g.s. of the final nucleus:
\bea
| IAS \rangle &=& %\frac{ \hat T^{-}}{\sqrt{2T_0}} |0_i^+\rangle
|T_0\, T_0-1\rangle_i + \sum_s\alpha_s\,|T_0-1 \, T_0-1\rangle_s , \label{IAStot}\\
| 0_f^+ \rangle & = &|T_0-2\, T_0-2\rangle_f + \sum_s\beta_s\,|T_0-1\, T_0-2\rangle_s +  
\gamma\, |T_0\, T_0-2\rangle_i . \label{0f_tot}
\eea 
It suffices for description of the IAS wave function Eq.~(\ref{IAStot}) to take into account only the admixtures of all $0^+$ states with the normal isospin $T_0-1$ in the intermediate nucleus. For the g.s. of the final nucleus, Eq.~(\ref{0f_tot}), the admixtures of all $T$-greater $0^+$ states must be considered as well as the only $\Delta T=2$ admixture of the ideal DIAS $|T_0\, T_0-2\rangle_i$ whose amplitude $\gamma$ is the quantity to be determined experimentally. The amplitudes $\alpha$ and $\beta$ of the $\Delta T=1$ isospin admixtures in Eq.~(\ref{IAStot},\ref{0f_tot}) are mainly determined by the mean Coulomb field $U_C$, while the amplitude $\gamma$ is essentially determined by the isotensor two-body Coulomb interaction.

In order to be able to determine $\gamma$ experimentally, one would need the reaction channel going through the admixture of the ideal DIAS:
$ \gamma\,|n\rangle \otimes | T_0\, T_0-2 \rangle_i \to |p\rangle \otimes |T_0\, T_0-1\rangle_i$, with  $T=T_0\pm{1\over2}, T_z=T_0-{3\over2}$, to dominate the cross section of the $(n,p)$ reaction %on the g.s. of the final nucleus 
at the zero scattering angle.
In fact, the $\Delta T=1$ isospin admixtures allow for other open competitive channels:
$ |n\rangle \otimes \beta| T_0-1\, T_0-2 \rangle \to |p\rangle \otimes |T_0\, T_0-1\rangle$ (with $T=T_0-{1\over2}$, channel I) and
$ |n\rangle \otimes | T_0-2\, T_0-2 \rangle \to |p\rangle \otimes \alpha |T_0-1\, T_0-1\rangle$ (with $T=T_0-{3\over2}$, channel II), whose relative contributions we are going to estimate.

Let us first estimate the cross section in the channel I. The corresponding T-matrix for the one-step reaction is:
\be
T^{\mathrm{(I)}}_{np} (0_f^+\to IAS) \propto \langle p|\otimes _i\langle T_0\, T_0-1| \hat V_{str} 
|n\rangle \otimes \sum_s \beta_s | T_0-1\, T_0-2 \rangle_s , \label{T1}
\ee
where T-greater states in the final nucleus are the analog states of $0^+$ states in the intermediate nucleus: $| T_0-1\, T_0-2 \rangle_s = \frac{ \hat T^{-}}{\sqrt{2T_0-2}}| T_0-1\, T_0-1 \rangle_s$, and $\hat V_{str}$ is the interaction between the projectile and the target which is assumed to be dominated by strong interaction. The admixture amplitudes $\beta_s$ in the first order of perturbation theory with respect to the Coulomb mean field read:
\be
\beta_s= \displaystyle \frac{_s\langle  T_0-1\, T_0-2 | \hat U_C | T_0-2\, T_0-2 \rangle_f}{E_{s_>}-E_{0_f^+}}
= \displaystyle \frac{_s\langle  T_0-1\, T_0-1 | \hat U_C^{+} | T_0-2\, T_0-2 \rangle_f}{\sqrt{2(T_0-1)}(E_{s}+\Delta_C-E_{0_f^+})},
\ee
where the charge-changing Coulomb mean field is defined as
$\hat U^{+}_C=[\hat T^{+},\hat U_C]= \frac{Ze^2}{2R}\displaystyle\sum_a (3-\frac{r_a^2}{R^2})\tau^+_a \ (r_a<R)$ inside the nucleus, and $\Delta_C$ is the Coulomb displacement energy. Then, by making use of isospin symmetry of $\hat V_{str}$ the expression (\ref{T1}) can further be transformed to acquire the form:
\be
T^{\mathrm{(I)}}_{np} (0_f^+\to IAS) \propto \displaystyle \sum_s
\frac{ _f\langle T_0-2\, T_0-2  | \hat U_C^{-} | T_0-1\, T_0-1 \rangle_s}
{\sqrt{2T_0}(E_{s}+\Delta_C-E_{0_f^+})}
{_s\langle} T_0-1\, T_0-1| a_n(k') \hat V_{str} a_p^\dagger(k)| T_0\, T_0 \rangle_i .
\label{T11}
\ee
The matrix element $_s\langle T_0-1\, T_0-1| a_n(k') \hat V_{str} a_p^\dagger(k)| T_0\, T_0 \rangle_i$ determines the T-matrix for the direct forward-scattering $(p,n)$ reaction on the g.s. of the initial nucleus with the excitation of monopole states in the intermediate nucleus, mainly those which form the isovector monopole resonance (IVMR). One would need a reaction code to calculate such a T-matrix, but for a rough estimate we assume here that this $(p,n)$ reaction cross section is proportional to the respective monopole strength of a $0^+$ component of the IVMR: 
\be
\sigma_{pn}(0_i^+\to IVMR_s)=\sigma_{0} \left|\langle IVMR_s | \hat R^{-} | 0_i^+ \rangle \right |^2,
\ee
where $\sigma_{0}$ is the unit cross section, and $\hat R^{-}=\sum_a \frac{r_a^2}{R^2}\,\tau^-_a$ is the monopole charge-changing operator. With this simplification one can get
the ratio of two cross sections: 
\be
\displaystyle 
r^{\mathrm{(I)}}\equiv\frac{\sigma^{\mathrm{(I)}}_{np}(0_f^+\to IAS)}{\sigma_{pn}(0_i^+\to IVMR)}=
\frac{1}{2T_0}\left(\frac{Ze^2}{2R}\right )^2 \displaystyle\frac{\displaystyle\left|\sum_s \frac{\langle 0_f^+ | \hat R^{-} | s \rangle \langle s | \hat R^{-} | 0_i^+ \rangle}{E_{s}+\Delta_C-E_{0_f^+}} \right |^2}{\displaystyle\sum_s \left| \langle s | \hat R^{-} | 0_i^+ \rangle \right |^2}.
\label{r1}
\ee
The nominator in this expression has a form 
%similar to a Fermi $2\nu\beta\beta$ 
of the squared amplitude of a double %charge-changing 
monopole transition between the g.s. of the initial and final nuclei, and the denominator contains the total $\beta^-$ monopole strength of the IVMR excited from the g.s. of the initial nucleus.

The same transformation with the T-matrix in the channel II
\be 
T^{\mathrm{(II)}}_{np} \propto
 _f\langle T_0-2\, T_0-2  | a_n(k')\,\hat V_{str} \, a_p^\dagger(k)  | T_0-1\, T_0-1 \rangle_s \displaystyle \frac{_s\langle T_0-1\, T_0-1| \hat V_C^{-} | T_0\, T_0 \rangle_i}
{\sqrt{2T_0}(E_{s}-\Delta_C-E_{0_i^+})}
\ee
leads to the second ratio:
\be
\displaystyle 
r^{\mathrm{(II)}}\equiv\frac{\sigma^{\mathrm{(II)}}_{np}(0_f^+\to IAS)}{\sigma_{np}(0_f^+\to IVMR)}=
\frac{1}{2T_0}\left(\frac{Ze^2}{2R}\right )^2 
\displaystyle\frac{\displaystyle\left|\sum_s \frac{\langle 0_f^+ | \hat R^{-} | s \rangle \langle s | \hat R^{-} | 0_i^+ \rangle}{E_{s}-\Delta_C-E_{0_i^+}} \right |^2}
{\displaystyle\sum_s \left| \langle s | \hat R^{+} | 0_f^+ \rangle \right |^2},
\label{r2}
\ee
where the denominator contains the total $\beta^+$ monopole strength of the IVMR excited from the g.s. of the final nucleus.

Taking as an estimate $\sigma_{np}(\gamma DIAS\to IAS)\approx 10^{-6}\sigma_{pn}(0_i^+\to IAS)$ for the $(n,p)$-reaction cross section at the zero scattering angle
% on the g.s. of the final nucleus 
which is due to the admixture of the DIAS, 
one arrives at the following ratios of the cross sections in question:
\bea
\frac{\sigma^{\mathrm{(I)}}_{np}(0_f^+\to IAS)}{\sigma_{np}(\gamma DIAS\to IAS)}&\approx& 10^{6}r^{\mathrm{(I)}}\cdot\frac{\sigma_{pn}(0_i^+\to IVMR)}{\sigma_{pn}(0_i^+\to IAS)}\\
\frac{\sigma^{\mathrm{(II)}}_{np}(0_f^+\to IAS)}{\sigma_{np}(\gamma DIAS\to IAS)}&\approx& 10^{6}r^{\mathrm{(II)}}\cdot\frac{\sigma_{np}(0_f^+\to IVMR)}{\sigma_{pn}(0_i^+\to IAS)}.
\eea

For a numerical estimate 
%As a testing case 
we choose $^{82}$Se as a representative medium-mass $\beta\beta$-decaying nucleus.
Both $\Delta N =0$ and $\Delta N =2$ particle-hole (p-h) excitations contribute to the monopole matrix elements (the total monopole strength is dominated by the contribution of $\Delta N =2$ p-h excitations). For a simultaneous description of both types of excitations a very large single-particle (s.p.) basis should be used in a nuclear model which at the same time should preserve isospin. 
To circumvent this difficulty, here two separate methods are used for the estimate:   
$\Delta N =2$ p-h excitations are treated in the independent Bogolyubov quasiparticle picture within a large, $7\hbar\omega$, s.p. space, whereas $\Delta N = 0$ p-h excitations
are described within the continuum-QRPA with zero-range Landau-Migdal forces.
The estimates of the contribution of the $\Delta N =2$ p-h excitations are:
$r^{\mathrm{(I)}}=6.6\cdot 10^{-7}$; $r^{\mathrm{(II)}}=1.3\cdot 10^{-5}$.
In the continuum-QRPA calculation of the contribution of the $\Delta N =0$ p-h excitations the following substitution is used:
$\hat R^{-} \to \hat R^{-} - a \hat T^{-}$ 
($a=\frac{\langle IAS| \hat R^{-}| 0_i \rangle}{\langle IAS|\hat T^{-} | 0_i \rangle}$)
which allows to subtract from the monopole matrix elements the contribution of a spurious admixture of the IAS in monopole states that is due to a incomplete isospin conservation in the model. 
The estimates of the contribution of these $\Delta N = 0$ p-h excitations are:
$r^{\mathrm{(I)}}=6.4\cdot 10^{-8}$;  $r^{\mathrm{(II)}}=2.2\cdot 10^{-6}$.
Taking into account the ratios $\displaystyle\frac{\sigma_{pn}(0_i^+\to IVMR)}{\sigma_{pn}(0_i^+\to IAS)} \sim 0.1$, $\displaystyle\frac{\sigma_{np}(0_f^+\to IVMR)}{\sigma_{pn}(0_i^+\to IAS)} \sim 0.01$~\cite{remco}
one arrives at the conclusion that $\displaystyle\frac{\sigma^{\mathrm{(I)}}_{np}(0_f^+\to IAS)}{\sigma_{np}(\gamma DIAS\to IAS)}\sim \frac{\sigma^{\mathrm{(II)}}_{np}(0_f^+\to IAS)}{\sigma_{np}(\gamma DIAS\to IAS)}\sim 0.1$. So, indeed the cross section $\sigma_{np}(0_f^+\to IAS)$  at the zero scattering angle is dominated by the admixture of the ideal DIAS in the g.s. of the final nucleus.

In the above analysis the assumption of a weak mixing between the IAS and the $0^+$ states
of normal isospin in the intermediate nucleus has been employed. However, in heavy nuclei the density of $0^+$ states in the vicinity of the IAS becomes so high that the IAS gets spread (the resonant mixing occurs). This effect is still to be taken into consideration. Nevertheless, in the case of the IAS of $^{48}$Ca ($T=4, T_z=3$) in $^{48}$Sc the weak-mixing assumption seems to be realized. The corresponding estimates will be published elsewhere.

To conclude, it has been shown in Ref.~\cite{Rod09} that the Fermi $0\nu\beta\beta$ nuclear matrix element can be reconstructed 
if one is able to measure the isospin-forbidden matrix element from the g.s. of the final nucleus to the isobaric analog state in the intermediate nucleus, for instance by means of charge-exchange reactions of the $(n,p)$-type. Here, simple estimates  have shown that indeed the tiny cross section $\sigma_{np}(0_f^+\to IAS)$ is dominated by the admixture of the DIAS in the g.s. of the final nucleus provided the mixing between the IAS and the $0^+$ states of normal isospin in the intermediate nucleus is weak and can be treated perturbatively. A measurement of such a cross section would definitely be a very difficult task, but it can advance a lot our knowledge of the $0\nu\beta\beta$ nuclear matrix elements.

The author acknowledges support of the DFG within the SFB TR27 ``Neutrinos and Beyond''.

\end{document}